\documentclass[fleqn,usenatbib]{mnras}

\usepackage{newtxtext,newtxmath}

\usepackage[T1]{fontenc}
\usepackage{ae,aecompl,ulem,color}

\usepackage{graphicx}	
\usepackage{amsmath}	
\usepackage{amssymb}	
\usepackage{embedfile} 
\embedfile{mnras_template.tex}

\newcommand {\delA}[1] {\textcolor{red}{}}

\title[Radio from unbound TDE debris]{Radio Emission from the unbound Debris of Tidal Disruption Events}

\author[A. Yalinewich et al.]{
A. Yalinewich$^{1}$,\thanks{E-mail: almog.yalin@gmail.com}
E. Steinberg$^{2}$,
T. Piran$^{3}$ and
J. H. Krolik$^4$
\\
$^{1}$Canadian Institute for Theoretical Astrophysics, 60 St. George St., Toronto, ON M5S 3H8, Canada\\
$^{2}$Columbia Astrophysics Laboratory, Columbia University, 550 West 120th Street, New York, NY 10027, USA\\
$^{3}$Racah Institute of Physics, the Hebrew University, 91904, Jerusalem, Israel\\
$^4$Department of Physics and Astronomy, Johns Hopkins University, Baltimore, MD 21228, USA
}

\date{Accepted XXX. Received YYY; in original form ZZZ}

\pubyear{2019}

\begin{document}
\label{firstpage}
\pagerange{\pageref{firstpage}--\pageref{lastpage}}
\maketitle

\begin{abstract}
When a star gets too close to a supermassive black hole, it is  torn apart by the tidal forces. Roughly half of the stellar mass becomes unbound  and flies away  at tremendous velocities - around $10^4$ km/s. In this work we explore the idea that the shock produced 
by the interaction of the unbound debris with  the ambient medium  gives rise to  the synchrotron radio emission observed in several TDEs. 
We use a moving mesh numerical simulation to study the evolution of the unbound debris and the bow shock around it. We find that as the periapse distance of the star decreases, the outflow becomes faster and wider.  A tidal disruption event whose periapse distance is a factor of 7 smaller than the tidal radius can account for the radio emission observed in ASASSN-14li. This model also allows us to obtain a more accurate estimate for the gas density around the centre of the host galaxy of ASASSN-14li.
\end{abstract}




\section{Introduction}


The observed radio spectra of several tidal disruption events are consistent with synchrotron emission.  The origin of this synchrotron emission is debated, with various models including a jet \citep{vanVelzen2016AASASSN-14li}, a disk wind driven by super Eddington outflow \citep{Alexander2016DISCOVERYASASSN-14li} or the unbound debris of the tidally disrupted star \citep[][denoted hereafter K16]{Krolik2016ASASSN-14li:EVENT}. Using high resolution numerical simulations  we explore the last possibility, in which
the radio signal arises from the collision of the unbound debris with gas surrounding the central black hole.

Two of the best-observed radio TDEs are  SWIFT J164449.3+573451 and ASASSN-14li (hereafter 14li) \citep{Holoien2015SixASAS-SN}. For SWIFT J164449.3+573451, the peak X-ray luminosity  far exceeds the Eddington luminosity of the central black hole, providing strong evidence for a relativistic jet, \citep[e.g.][]{Bloom2011AStar}. \cite{Zauderer2013RADIOEMISSION} suggested that the radio emission arose from the interaction of this jet with the surrounding matter.  
The other event, 14li, did show signs for a jet, but  much less energetic than in the case of SWIFT J164449.3+573451 \citep{Kara2018UltrafastASASSN-14li}. Investigations of 14li focused on models where the outflow originates from the accretion disc formed by the bound debris, namely a jet \citep{vanVelzen2016AASASSN-14li} or a quasi- spherical wind driven by a super Eddington accretion disc \citet{Alexander2016DISCOVERYASASSN-14li}. On the other hand \citet[][denoted hereafter K16]{Krolik2016ASASSN-14li:EVENT} found using equipartition analysis that the radio emitting regions expand at a velocity comparable  to that of the unbound debris; on this basis, they suggested that the unbound debris are the source of the observed radio signal. Moreover, because the total energy of the unbound debris is similar to the energy injected into a supernova remnant, one would expect the same mechanism that produces radio emission in supernova remnants to also operate in the case of tidal disruption events \citep{Guillochon2016UNBOUNDHOLES}.

The observational situation for ASASSN-14li is further complicated by a
possible correlation between the radio and X-ray lightcurves \citep{Pasham2018DiscoveryCoupling}.  Taken at face value, this poses difficulties for
both jet and ejecta models for the radio emission: for the former
because a synchrotron self-Compton spectrum \citep[as envisioned by][]{Pasham2018DiscoveryCoupling} is inconsistent with the observed X-ray spectrum
\citep[extremely soft, fit by a thermal spectrum with kT ~50 eV][]{Miller2015FlowsHole}, for the latter because the ejecta are not expected to produce
X-rays at all.  However, the global significance of this correlation is
somewhat unclear, as it relates ~10\% modulations of two lightcurves that
have been de - trended by different power-laws. Here we will compare our
predictions to 14li because of the quality of the available data, but we
emphasise that the general properties we discuss should be more widely
applicable.

In this work we explore further the idea that the radio emission arises from the unbound debris. We show that for a given synchrotron source the solid angle subtended by the source is strongly related to its  velocity  (see \ref{sec:synchrotron}). Specifically, if  the solid angle subtended by the emitting material is too small, the matter must be moving relativistically to match the observations. The questions we  ask here are: First, what is the solid angle subtended by the outgoing debris? Second, can this opening angle and the theoretical velocity explain the observations? These two questions hold the key to the validity of the model, as we show in section \ref{sec:synchrotron}.

The bulk of the unbound debris subtends a very small solid angle. We recall from the discussion above that to reproduce the observed radio signal, outflows with small opening angles must be moving at relativistic velocity. This may seem, at first, to be inconsistent with the estimated non-relativistic velocity of the ejecta. However, a tiny fraction of the disrupted mass, less than $10^{-3} M_{\odot}$, is sufficient to power the radio emission (K16). Previous work has shown that a small fraction of the unbound debris moving ahead of the stream expands more than the bulk \citep{Rosswog2009TIDALHOLES}. Unfortunately, so far the properties of this fastest ejecta have not been characterised. Moreover, for reasons discussed  below, it would have been difficult to obtain reliable results using a simulation with a static mesh or with a SPH code. We have therefore carried out numerical simulation of a tidal disruption event with a moving mesh code that captures the dynamics of the fastest ejecta, determining its mass and its range of opening angles. We then characterise the resulting radio signature.

The plan of the paper is as follows. In section \ref{sec:obs} we briefly describe the radio observations of 14li. We recount some of the results concerning synchrotron emission, the unbound debris and bow shocks in section  \ref{sec:bg}. In section \ref{sec:sim} we discuss our numerical simulation. In section \ref{sec:14li} we discuss 14li and its radio signal. Finally, in section \ref{sec:conclusion} we discuss the results and their implications.

\section{Radio Observations of 14li}
\label{sec:obs}

TDE 14li was the nearest TDE observed so far, and had a bright radio signature. It was observed in radio for the first eight months after its detection using the Arcminute Microkelvin Imager at 15.7 GHz  and the Westerbork Synthesis Radio Telescope at 1.4 GHz \citep{vanVelzen2016AASASSN-14li}, and with the Karl G. Jansky Very Large Array in a range of frequencies between 1 and 25 GHz \citep{Alexander2016DISCOVERYASASSN-14li}. Over the course of this period, the peak flux decreased from about 1.8 to 0.6 mJy, while the peak frequency decreased from about 2 to 0.2 GHz. Later, up to two years after the peak of the 14li flare, Arcminute Microkelvin Imager observations at 15.7 GHz showed that the flux plateaus at 244 $\mu \rm Jy$. This level  is consistent with the quiescent radio emission that preceeded the TDE  \citep{Bright2018Long-termASASSN-14li}. Finally, \citet{Romero-Canizales2016TheEVN} used the European Very Long Baseline Interferometry Network (EVN) to resolve the structure around 14li. The observations revealed a collimated outflow, but they were unable to determine whether or not the outflow is moving at relativistic speeds.

\section{Background} \label{sec:bg}
\subsection{Synchrotron Emission} \label{sec:synchrotron}

Consider an outflow that interacts  with surrounding matter whose electron number density  is $n(r)$. The edge of the outflow moves at a velocity $v$ at a distance $r$ from the source. The thermal energy density of the shocked region is roughly given by $m_p n v^2$, where  $m_p$ is the proton mass. Assuming a fraction $\varepsilon_b$ of the thermal energy is converted into magnetic field, we obtain a magnetic field strength
\begin{equation}
B \approx \sqrt{\varepsilon_b m_p n } v \, .
\end{equation}
Similarly, we assume that a fraction, $\varepsilon_e$, of the thermal energy goes into accelerating supra-thermal electrons. We further assume that a fraction $\chi \ll 1$ of those are accelerated to relativistic energies and that the energy distribution is a power-law, so
\begin{equation}
\frac{d n}{d \gamma} \approx \chi n \left(\frac{\gamma}{\gamma_0}\right)^{-p} \frac{1}{\gamma_0} \quad \quad {\rm for} \quad \gamma>\gamma_0\approx \frac{\varepsilon_e}{\chi} \frac{m_p}{m_e}\frac{v^2}{c^2} \ .
\end{equation}
We assume that $p>2$ so that the energy is dominated by the lower energy electrons with $\gamma\approx \gamma_0$. We are interested in the case where the outflow propagates with constant opening angles in the plane of motion of the star around the black hole $\theta_{y}$ and normal to the plane of motion $\theta_{z}$ (see figure \ref{fig:sketch}). The volume of the radiating region is then $\simeq \theta_{z} \theta_{y} r^3$. The characteristic synchrotron frequency emitted by an electron with Lorentz factor $\gamma$ is $\omega_s \approx \frac{q B}{m_e c} \gamma^2$, where $q$ is the elementary charge.  The population-integrated optically thin spectral luminosity is \citep{Rybicki1979RadiativeAstrophysics}
\begin{equation}
L_{\omega, \rm thin} \approx \left[c^{- 2 p + 3} r_{e}^{\frac{p+5}{4}} m_{e}^{ \frac{-5 p+7 }{4} } m_{p}^{\frac{5 p-3}{4} }\right]  \times 
\end{equation}
\begin{equation*}
\theta_{z} \theta_{y} r^3 \chi^{- p + 2} \epsilon_{b}^{\frac{p+1}{4} } \epsilon_{e}^{p - 1} \omega^{- \frac{p-1}{2}} v^{\frac{5 p-3 }{2}} n^{\frac{p+5}{4} } \,
\end{equation*}
where $r_e$ is the  classical electron radius  and we used square brackets to cluster together the physical constants. If the system is optically thick, the luminosity is
\begin{equation}
L_{\omega, \rm thick} \approx \left[\frac{m_{e}^{\frac{5}{4}}}{\sqrt[4]{r_{e}} \sqrt[4]{m_{p}}}\right] \theta_{z} \theta_{y} r^2 \frac{ \omega^{\frac{5}{2}}}{\sqrt[4]{\epsilon_{b}}  \sqrt[4]{n} \sqrt{v}} \, .
\end{equation}
The transition between optically thin and thick occurs at a frequency 
\begin{equation}
\omega_{\rm sa} \approx \left[ c^{\frac{2 \left(- 2 p + 3\right)}{p + 4}} r_e^{\frac{1}{2}\frac{p + 6}{p + 4}} m_{e}^{\frac{- 5 p + 2}{2 \left(p + 4\right)}} m_p^{\frac{1}{2}\frac{5 p - 2}{p + 4}} \right]  \times \label{eq:break_frequency}
\end{equation}
\begin{equation*}
\left(\theta_{y} r\right)^{\frac{2}{p + 4}} \chi^{\frac{2 \left(- p + 2\right)}{p + 4}} \epsilon_{b}^{\frac{p + 2}{2 \left(p + 4\right)}} \epsilon_{e}^{\frac{2 \left(p - 1\right)}{p + 4}} v^{\frac{5 p - 2}{p + 4}} n^{\frac{1}{2}\frac{p + 6}{p + 4}} \, .
\end{equation*}
The spectral luminosity at the break frequency is
\begin{equation}
L_{\omega, \rm sa} \approx \left[m_p^{\frac{1}{2} \frac{12 p - 7}{p + 4}} r_{e}^{\frac{1}{2}\frac{2 p + 13}{p + 4}} \left(c^{5} m_{e}^{\frac{5}{2}}\right)^{\frac{- 2 p + 3}{p + 4}}\right]  \times \label{eq:break_luminosity}
\end{equation}
\begin{equation*}
\theta_{z} \theta_{y}^{\frac{p+9}{p+4}} \chi^{\frac{5 \left(- p + 2\right)}{p + 4}} \epsilon_{b}^{\frac{2 p + 3}{2 \left(p + 4\right)}} \epsilon_{e}^{\frac{5 \left(p - 1\right)}{p + 4}} v^{\frac{12 p - 7}{p + 4}} \left(r \sqrt{n} \right)^{\frac{2 p + 13}{p + 4}} \, .
\end{equation*}
Eliminating the density $n$ and substituting $r= t v$, where $t$ is the time since periapse passage, we find that the velocity is given by
\begin{eqnarray}
v & \approx& \left\{  \left[c^{2 p - 3}  m_e^{- 8} m_p^{-p+2} \right] \left[\chi^{p - 2} \epsilon_{b} \epsilon_{e}^{- p + 1}\right] \right.  \,\\ 
 &    \times& \left. \left[\theta_{y}^{- p - 6} \theta_{z}^{- p - 7}\right] \left[L_{\omega, \rm sa}^{p + 6} \omega_{\rm sa}^{- 2 p - 13} t^{- 2 p - 13}\right] \right\}^{\frac{1}{4p + 9}} \, . \nonumber
\end{eqnarray}
Due to the size of the expression, we clustered together physical constants, parameters related to Fermi acceleration, opening angles and measured quantities, in that order. For a typical value of the power-law slope $p=2.5$, the velocity scales with the opening angles as $v \propto \theta_y^{-0.44} \theta_z^{-0.5}.$
For a given luminosity and frequency, smaller opening angles require a larger velocity; conversely, lower break frequency and higher luminosity at the break imply higher velocity for fixed opening angles.
Put differently we find that there is a degeneracy in the synchrotron solution. We can obtain the same signal with different conditions at the emitting regions. A region with a smaller angular size requires a larger radius  and this requires, in turn, a larger velocity.

The number density of the external gas associated with the observed luminosity and break frequency is a function of time, which, in this constant-speed model, is proportional to distance from the black hole:
\begin{eqnarray}
n &\approx& \left[ r_e^{-1} c^{\frac{6 \left(2 p - 3\right)}{4 p + 9}}  m_{e}^{\frac{20 p-3}{4 p + 9}} m_{p}^{\frac{3-10 p}{9+4 p}}\right]  
    \left[t^{\frac{2 \left(10 p - 3\right)}{4 p + 9}} L_{\omega, \rm sa}^{-\frac{10 p}{4 p + 9}}  \omega_{\rm sa}^{\frac{4 \left(7 p + 3\right)}{4 p + 9}} \right] \nonumber\\
&   \times& ~  
\left[\theta_{y}^{\frac{10 p}{4 p + 9}} \theta_{z}^{\frac{2 \left(5 p -3\right)}{4 p + 9}}    \chi^{\frac{6 \left(p - 2\right)}{4 p + 9}} \epsilon_{b}^{-\frac{4 p + 3}{4 p+9}} \right ]  \left[\epsilon_{e}^{-\frac{p-1}{4p + 9}}\right] \ . 
\end{eqnarray}
Due to the length of these expressions, we've clustered terms so that physical constants appear in the first block, directly measured quantities appear in the second block, opening angles in the third block and quantities related to Fermi acceleration in the fourth block. For a typical value $p = 2.5$, the density scales with the opening angle as $n \propto \theta_y^{1.3} \theta_z^{1}$. Hence, for fixed other parameters, the density decreases as the opening angle decreases. Smaller opening angle implies a smaller external density at the same observation time, but the radius at which this density is found depends on the velocity.  It is possible that at the same distance the inferred density might increase with decreasing opening angle, depending on the spatial density profile.

These  relations 
are found in fact already in the estimates of \cite{Alexander2016DISCOVERYASASSN-14li} and K16. Both find that 
the velocity is constant, which is consistent with the requirement that only a small amount of matter   (unbound debris for K16, a weak wind for \cite{Alexander2016DISCOVERYASASSN-14li}) collides with the external matter. Both also find that the velocity obtained depends on the assumed opening angles, with larger velocities for smaller opening angles. 

Note that K16 used a different formalism than is presented here. Instead of using opening angles, they described the outflow in terms of an areal filling factor $f_A$ and a volume filling factor $f_V$: if $r$ is the radius of the outflow, the surface area of the emitting region is $f_A r^2$ ($f_A = 4$ in the isotropic case) and $f_V r^3$ is the volume ($f_V = 4 \upi/3$ for the isotropic case). The conversion between our formalism and theirs is given by $f_A \approx \theta_y \theta_z$ and, if the emitting region is radially thick, $f_V \approx f_A$. 

\subsection{The Unbound Debris}

We consider a tidal disruption of a star of mass $M_*$ by a central black hole of mass $M_{BH} = Q M_*$. Such a star would be disrupted if its distance from the black hole drops below the tidal radius
\begin{equation}
R_t \approx R_* Q^{1/3}\ , 
\end{equation}
where $R_*$ is the stellar radius. As a result of the disruption, roughly half of the debris remains bound to the black hole, while the rest flies away from the black hole.
The characteristic velocity at infinity of the bulk of the unbound debris is  (e.g. K16)
\begin{equation}
v_{\infty} \approx 9,000 ~{\rm km/sec}~\left(\frac{k/f}{0.3}\right)^{-1/6} \left(\frac{M_*}{M_{\odot}}\right)^{1/2} \left(\frac{R_*}{R_{\odot}}\right)^{-1/2} \left(\frac{Q}{10^6}\right)^{1/6} \, 
\end{equation}
where the factor $k/f$ characterises the internal structure of the star; for a fully convective star, it is $\simeq 0.03$ \citep{Phinney1989ManifestationsCenter}, and this is the appropriate match to the polytrope we will simulate.

The bulk of the ejecta rushes out within a narrow wedge whose opening angle in the plane of motion is $Q^{-1/6}$. The angle perpendicular to the plane of motion is $\sim Q^{-1/3}$ \citep{Strubbe2009OpticalHoles}. For a typical event with a mass ratio $Q \approx 10^{6}$, the solid angle $\Omega \approx Q^{-1/2} \approx 10^{-3}$.
Analysis of the synchrotron emission (see section \ref{sec:synchrotron} below) suggests that such a small angle would require either a relativistic velocity or an extremely large matter density to explain the observed signal. 
However, as only a small fraction of the matter is sufficient to produce the resulting radio emission (see K16) it is possible that the fastest material has a wider solid angle. Furthermore,  as we show below, the bow shock around the expanding wedge is rather wider than the debris wedge itself.

\subsection{The Bow Shock} \label{sec:bow_Shock}

Once the unbound outflow reaches a distance large enough that its kinetic energy is much larger in magnitude than its potential energy, it propagates at an almost constant velocity $v$. The angular size it subtends is determined by conditions during the disruption; we label the opening angle along the short axis $\theta_{z}' \ll 1$ (perpendicular to the plane of motion of the star around the black hole), and the opening angle along the long axis $\theta_{y}' \ll 1$ (parallel to the plane of motion). As will be discussed later on in this section, the opening angles of the emitting region are larger than the opening angles of the outflow, and that is why the latter are primed.

\begin{figure*}
\includegraphics[width=0.8\linewidth]{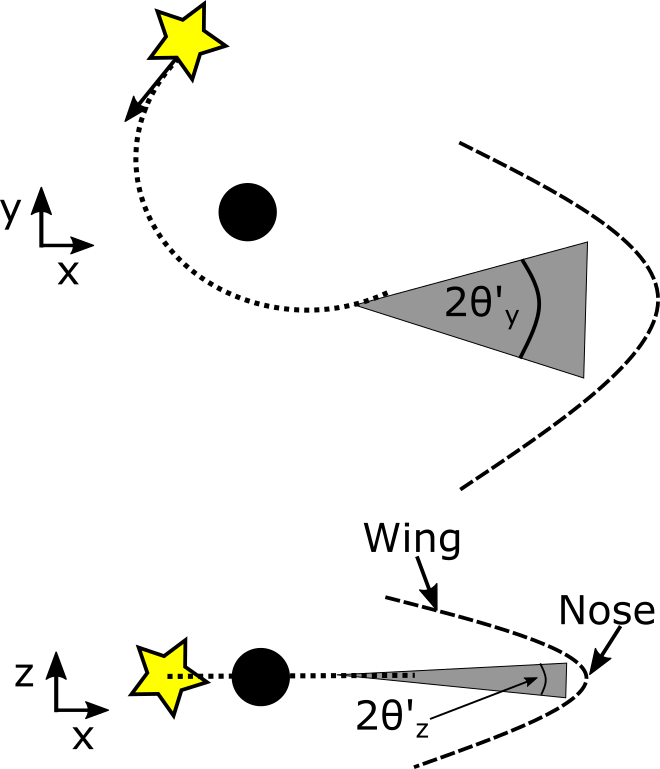}
\caption{A schematic description of the bow shock around the unbound debris (not to scale). The star before the disruption is represented by a yellow star, with an arrow denoting its direction of motion. The black hole is represented as a black circle. The orbital trajectory of the star's centre of mass is represented by a dotted curve. The unbound debris are represented by a grey wedge. The bow shock around the outflow (sometimes referred to as obstacle in the text) is represented by a dashed curve. We've also indicated the different opening angles of the outflow (marked with an apostrophe). As is discussed later on, the opening angles of the emitting region are larger by a factor of about 3 from those of the outflow (not shown).}
\label{fig:sketch}
\end{figure*}

As the outflow propagates outwards, it interacts with the surrounding matter and a bow shock forms around it. This bow shock covers a larger area then the outflow itself. We distinguish between two  different parts  (see figure \ref{fig:sketch}).  The ``nose" is the region  at the head of the bow shock  where the shock front is perpendicular to the incident velocity. The ``wings" extend 
along the outflow, and there the angle between the incident velocity and the shock is small and the shock is weaker.  We focus here on the emission from the nose.
The emission from the wings is less certain as it depends critically on the decay of the magnetic field from the nose to the wings (see appendix \ref{app:wind_emission} and section \ref{sec:14li}).

Simulations of bow shocks around spheres \citep{Yalinewich2016ASYMPTOTICNUMBER} and  slabs (see appendix \ref{app:cyl_bow_shock}) show that in each dimension the bow shock nose is three to four times larger than the obstacle. Thus,  we expect the effective opening angle of the shocked region to be larger by an order of magnitude than that of the outflow. We denote the opening angle of the shocked material along the short axis by $\theta_{z} \approx 3 \theta_{z}'$, and along the long axis by $\theta_{y} \approx 3 \theta_{y}'$. In other words, the opening angles of the emitting region are larger by about a factor of three compared with the opening angles of the outflow.

The calculations of the bow shock reported in \citet{Yalinewich2016ASYMPTOTICNUMBER}
used an obstacle of a constant size, whereas here the outflow has a constant opening angle, so the effective size of the barrier increases with time. 
However, we can approximate the solution using the steady state result, assuming that at every moment the bow shock is similar to the steady state solution with the current obstacle size.

\section{Simulations } \label{sec:sim}

\subsection{The Moving Mesh Simulation}

Tidal disruption events have previously been simulated mostly using either smooth particle hydrodynamics \citep[e.g.][]{Ayal2000TidalHole,  Rosswog2009TIDALHOLES, Hayasaki2016CircularizationHoles, Sadowski2016MagnetohydrodynamicalRelativity, Coughlin2017CircumbinaryEvents} or finite volume fixed mesh codes \citep[e.g.][]{ Guillochon2009THREE-DIMENSIONALBREAKOUT, Cheng2013RelativisticCoordinates, Mockler2018WeighingEvents, Curd2018GRRMHDDetectability}. 
Both approaches are appropriate for modelling certain aspects of tidal disruption events, but we argue that in this particular case they are inadequate. This is because we want to model a tiny fraction of the stellar mass, in a scenario that involves high levels of compression and numerous length scales.  Smooth particle hydrodynamics have difficulties in capturing shock waves \citep{Springel2010iEMesh} and are limited in mass resolution by the total mass of the simulation particles found within a smoothing length. Finite volume methods have a hard time handling sharp transitions in density, e.g., between the dense material inside the star and the vacuum outside. Numerical diffusion can cause a non-physical transfer of matter from the dense interior to the tenuous exterior, and this can contaminate the properties of the fastest component.

To circumvent these problems, we employ a moving mesh code \citep{Yalinewich2015Rich:Mesh}. While it is not perfectly Lagrangian, this code automatically adapts the size and position of the cells to the regions of interest in the calculations. At the same time, it can impose conservation conditions to an accuracy comparable to that of finite-volume grid codes by solving the same sort of Riemann problem at cell boundaries.   Hence, this code enjoys some of the benefits of both worlds. It can resolve shock waves while moving the computational cells along with the flow to minimise numerical diffusion.  On the other hand, because the code is not completely Lagrangian, it does not suffer from tangling of the numerical grid. Thus, the code is particularly well-suited to resolve the dynamics of  the small fraction of the outflow that spreads and propagates at high velocities.

\subsection{Setup}

We ran simulations of tidal disruption events with mass ratio $Q = 10^6$ and impact parameters $\beta=1$ and $\beta=7$ using the 3D version of the moving mesh RICH code \citep{Yalinewich2015Rich:Mesh}. We chose $\beta = 1$ to represent a shallow TDE, and $\beta = 7$ to represent a TDE where the star penetrates deep into the tidal radius, following a convention used in previous works \citep[e.g.][]{Guillochon2009THREE-DIMENSIONALBREAKOUT}.
We used $10^6$ mesh generating points to describe the star and $5 \cdot 10^5$ mesh generating points for the space outside it, which was filled with gas having density $10^{-24}\times$ the density at the star's centre. Throughout the simulation we use adaptive mesh refinement to guarantee that the mass of a single cell containing stellar material is between $10^{-6} M_{\odot}$ and $10^{-7} M_{\odot}$. The size of the computational domain was initially set to five stellar radii, but we increase it by 15\% every time the stellar material is about to reach the edge, and fill in the new volume with low density gas. The star was initialised as a polytrope with index $1.5$ at a distance of 3 times the tidal radius. For the equation of state, we use an ideal gas with an adiabatic index $5/3$.  Self-gravity is calculated with a kd-tree code including quadrupole terms. The simulations ran to time $t \approx 1.25 \cdot 10^6$ seconds after periapse passage (roughly 125 internal dynamical times of the star or, equivalently, Keplerian times at the Hills radius). At this stage  the distance between the star's centre of mass and the SMBH was $\approx 1.4\cdot 10^2$ times the tidal radius. From that point on we assume that fluid elements move along Keplerian orbits. Note that whilst the resolution is sufficient to determine the detailed behaviour of the small fraction of the outflow moving at the highest velocity, it is impractical to use the same simulation for the bow shock. This is because the bow shock is larger than the ejecta and takes longer then a tidal time to settle into a steady state. For this reason, the bow shock was simulated with a different setup, as is described in section \ref{sec:bow_Shock}.

\subsection{Results}

\begin{figure*}
\includegraphics[width=0.9\linewidth]{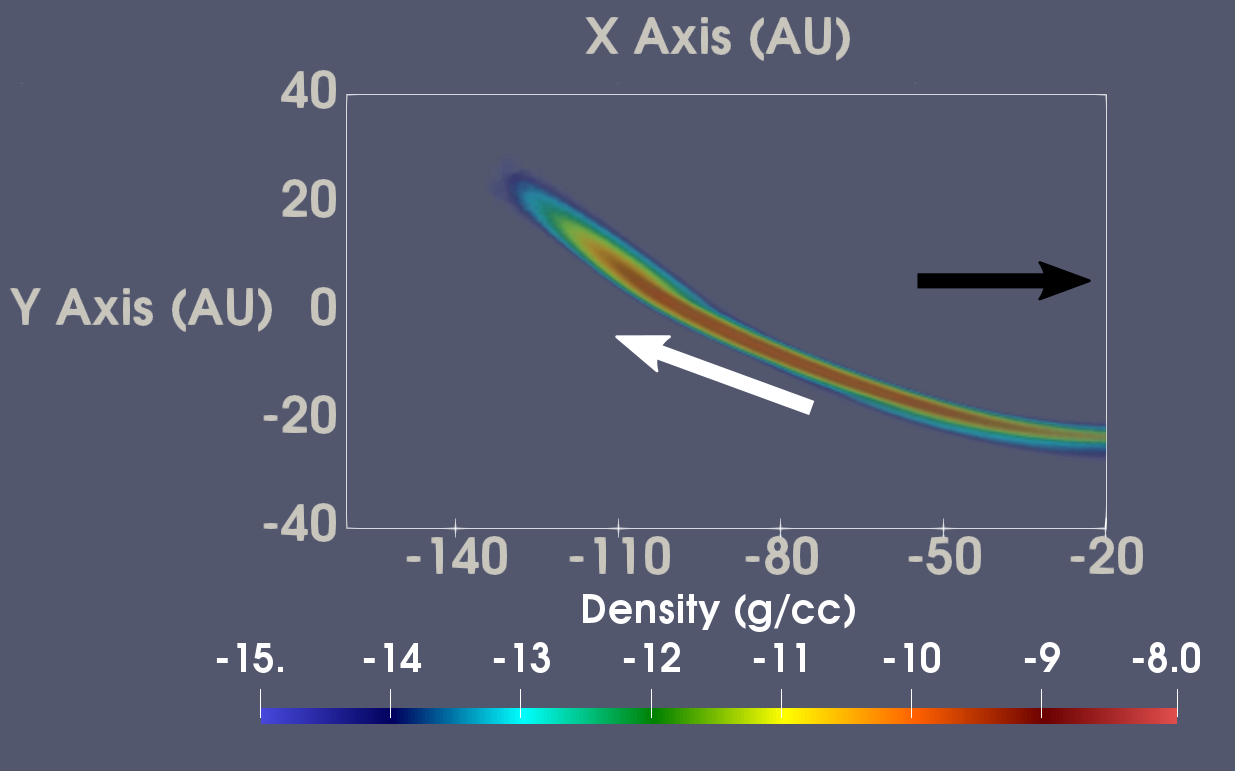}
\includegraphics[width=0.9\linewidth]{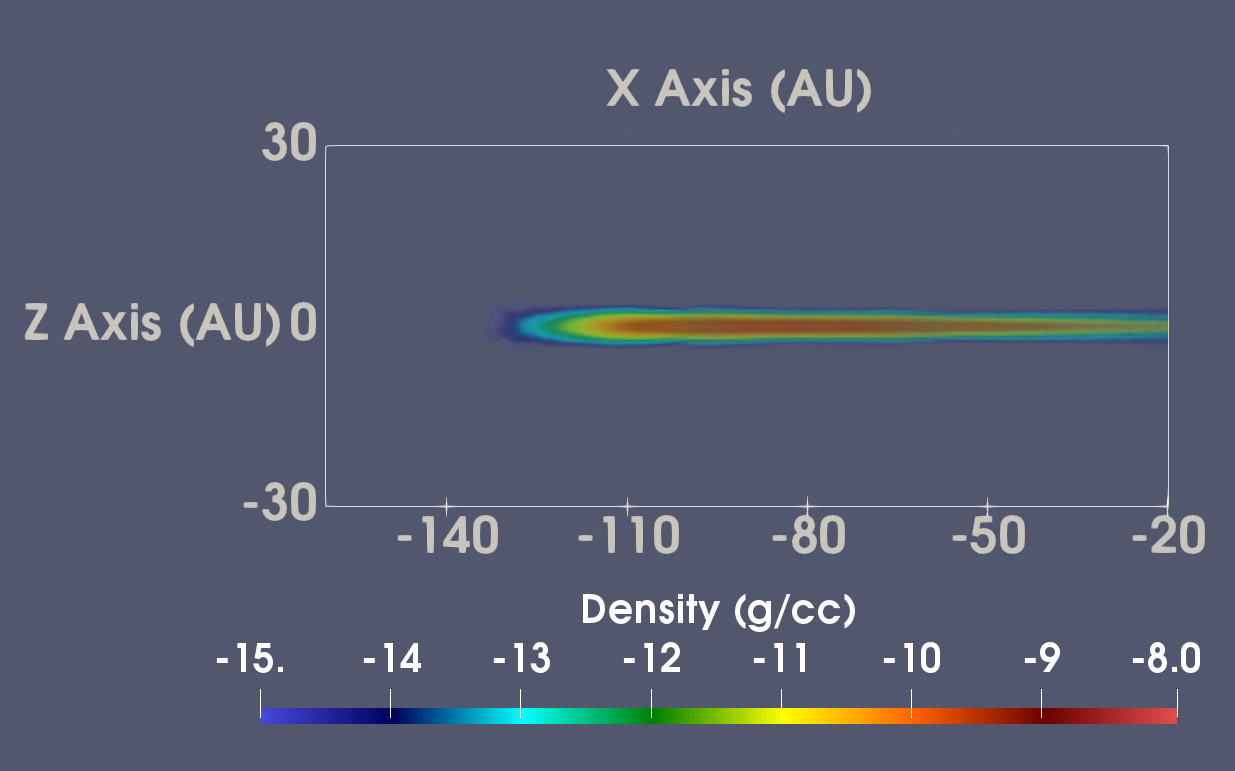}
\caption{Cross sectional view of the density distribution of the unbound material in the simulation of a TDE with $\beta=1$. The top panel shows a vantage point above the plane of motion, and the viewing direction of the bottom panel is in the plane of motion. The origin of the axes is the centre of the black hole. The white arrow indicates the velocity direction of the centre of mass of the unbound debris, and the black arrow points to the black hole. We note that this figure only shows the outflow, and not the bow shock around it.}
\label{fig:density_contours_beta_1}
\end{figure*}

\begin{figure*}
\includegraphics[width=\linewidth]{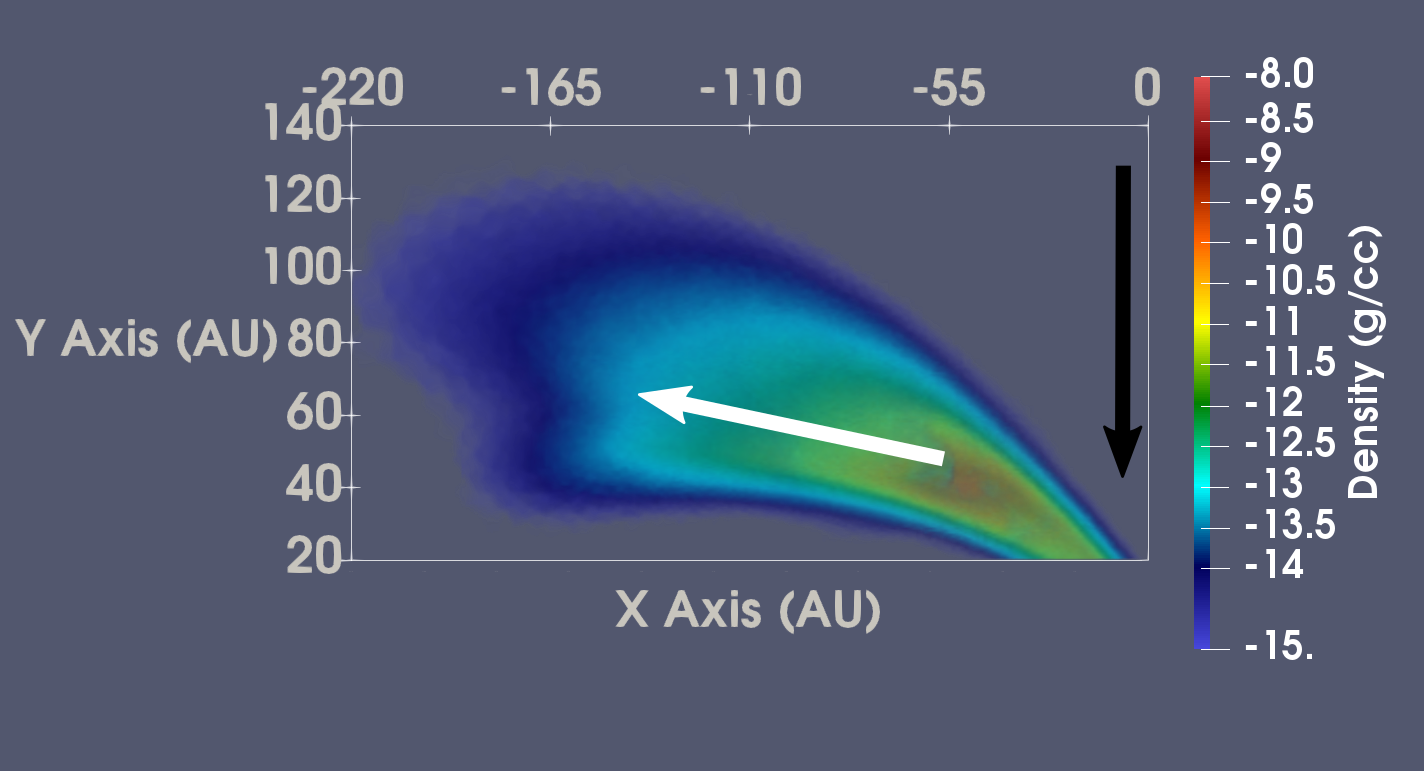}
\includegraphics[width=\linewidth]{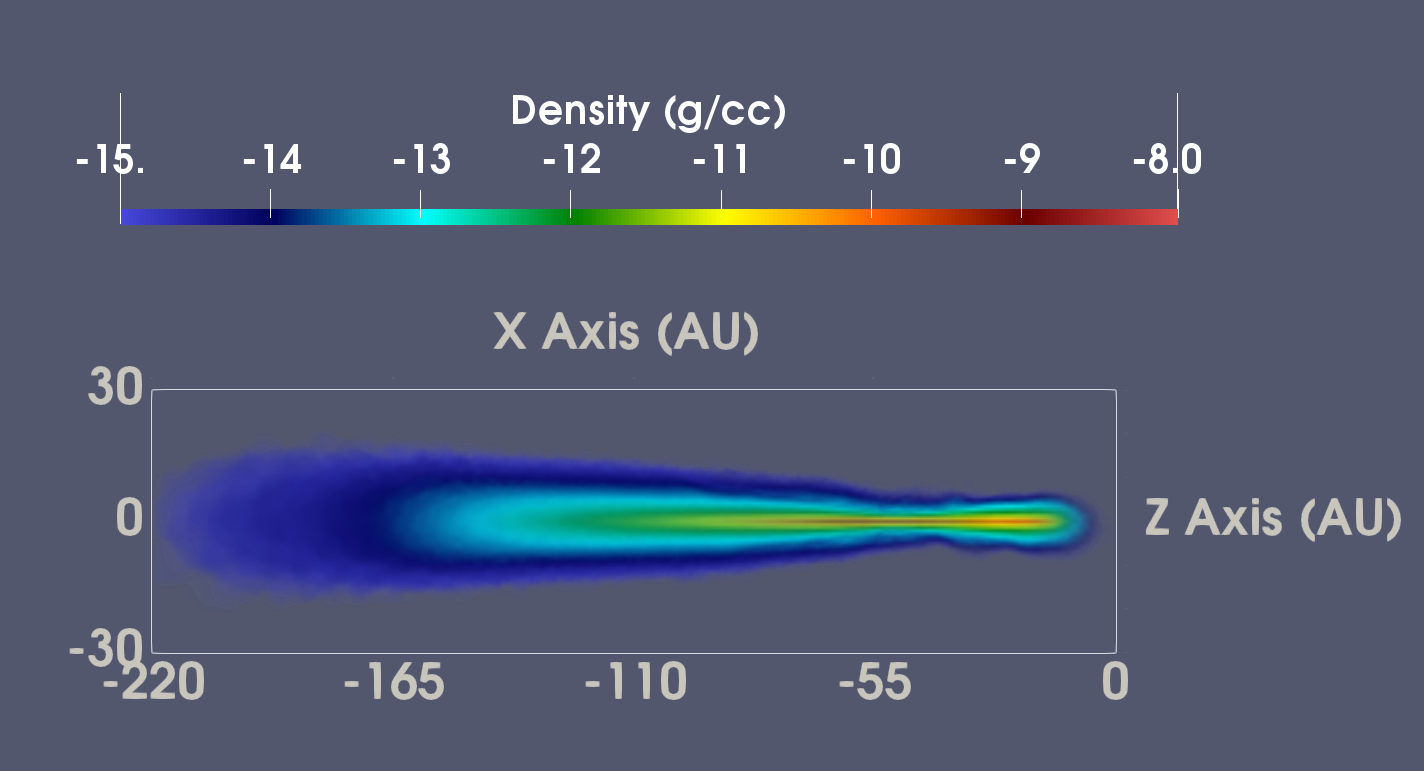}
\caption{Cross sectional view of the density distribution of the unbound material in the simulation of a tidal disruption event with $\beta=7$. The top panel shows a vantage point above the plane of motion, and the viewing direction of the bottom panel is in the plane of motion. The white arrow indicates the velocity direction of the centre of mass of the unbound debris, and the black arrow points to the black hole. We note that this figure only shows the outflow, and not the bow shock around it.}

\label{fig:density_contours_beta_7}
\end{figure*}

We present volume rendering of the density of the unbound debris for the simulation with $\beta = 1$ in figure \ref{fig:density_contours_beta_1}, and for $\beta = 7$ in figure \ref{fig:density_contours_beta_7}. In all cases the leading edge of the debris stream expands faster and wider than the bulk of the ejecta, as was seen in previous works. We note that the debris in the $\beta = 7$ case is much more spread out than in the $\beta = 1$ case.

\begin{figure}
	\includegraphics[width=0.9\columnwidth]{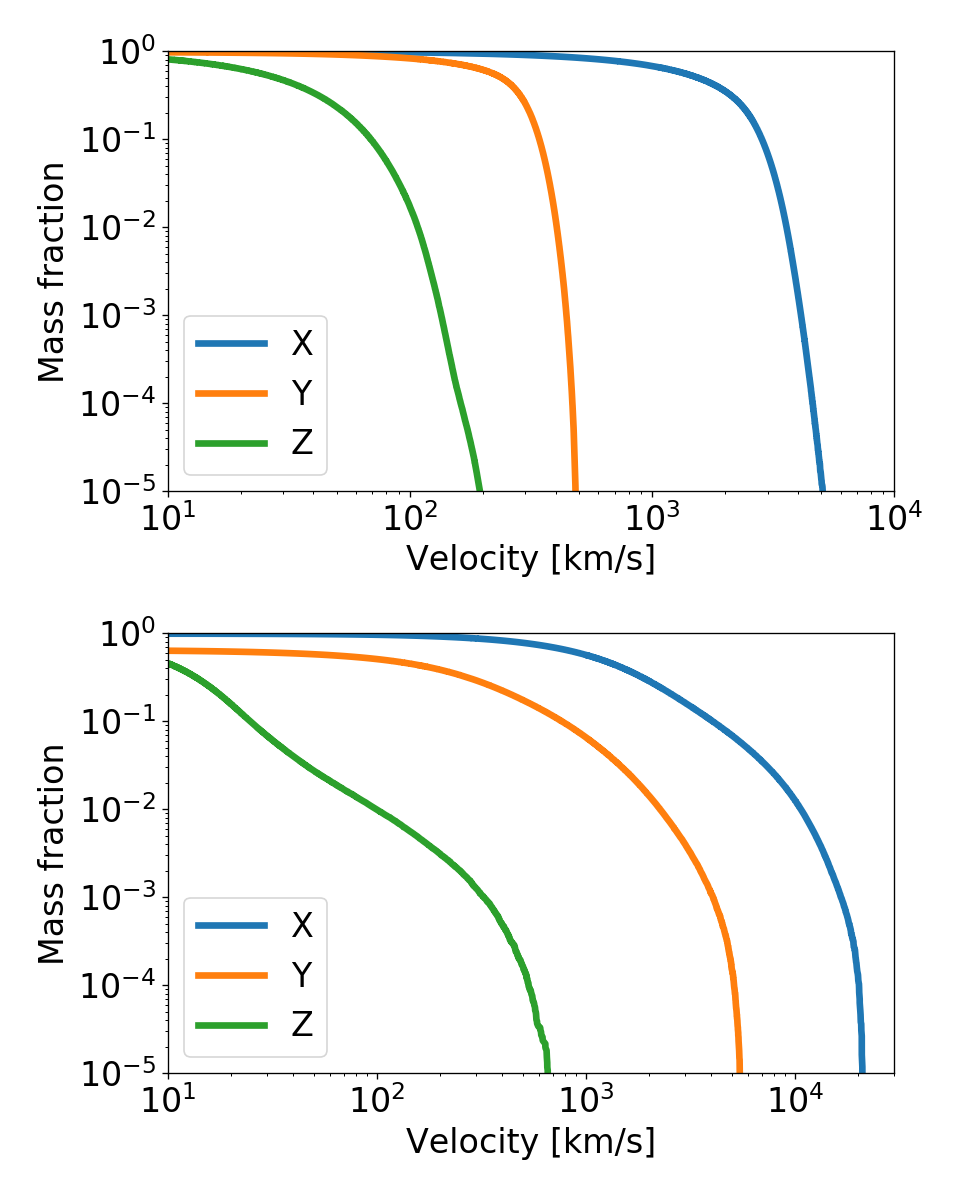}
    \caption{Cumulative distribution of the components of the velocity of the unbound debris, for $\beta=1$ (top) and $\beta=7$ (bottom). The components are aligned according to the principal directions of motion: Z is normal to the plane of motion, X is the direction of motion of the centre of mass of the unbound debris, and Y is in the plane of motion, but normal to X. This histogram only shows the magnitude of the velocity components, which are evaluated at an infinite distance from the black hole.}
    \label{fig:mass_velocity_histogram}
\end{figure}

The debris mass distribution permits construction of a principal axis system.  Figure \ref{fig:mass_velocity_histogram} depicts the cumulative distribution of mass with respect to velocity along each of the three axes of this system, i.e., $M(>v_i)/M_{\rm tot}$, where the subscript $i$ labels the axes. The ratios between the different components give the opening angles of the debris.

Figure \ref{fig:mass_velocity_histogram} shows that, in accordance with analytic estimates, most of the ejecta propagates as a thin wedge. However, a small fraction of the material expands at larger velocities and across a wider wedge. K16 estimated that even a fraction as little as $10^{-4}$ would suffice to produce the radio signal. In this work we consider a higher fraction $10^{-3}$ as a factor of safety, so we pay particular attention to this location in the cumulative mass distribution. 
For  $\beta=1$ the opening angles of this material are $\theta_{y} = 2 \tan^{-1} \left(0.1\right) \approx 0.2$ in the plane of motion and $\theta_{z} = 2 \tan^{-1} \left(0.02 \right) \approx 0.04$ perpendicular to the plane of motion (both estimated using figure \ref{fig:mass_velocity_histogram}). The corresponding velocity along the direction of the centre of mass of the unbound ejecta is $v_x \approx 5000 \, \rm km/s$, the velocity in the plane of motion perpendicular to $x$ is $v_y \approx 500 \, \rm km/s$, and perpendicular to the plane of motion the velocity is $v_z \approx 150 \, \rm km/s$. 
For the case of $\beta = 7$, the opening angles are $\theta_{y} = 2 \tan^{-1} \left(0.5\right) \approx 0.9$ in the plane of motion and $\theta_{z} = 2 \tan^{-1} \left(0.05\right) \approx 0.1$ perpendicular to the plane of motion. The velocity along the direction of the centre of mass of the unbound ejecta is $v_x \approx 2 \cdot 10^4 \, \rm km/s$, the velocity in the plane of motion perpendicular to $x$ is $v_y \approx 5000 \, \rm km/s$, and perpendicular to the plane of motion the velocity is $v_z \approx 400 \, \rm km/s$. 

\section{Application to ASASSN-14li} \label{sec:14li}

In section \ref{sec:sim} we presented numerical simulations and obtained the opening angles of the fastest and widest part of the outflow. In section \ref{sec:bow_Shock} we showed that a bow shock will develop around the outflow, and that its effective size would be about a factor of 3--4 larger then the outflow in both transverse directions. For  $\beta = 1$ we find that the shocked material subtends a solid angle of $\Omega  \approx 0.1$ sr, whereas for $\beta = 7$ we find $\Omega \approx 1 \, \rm sr$, each an order of magnitude larger than the corresponding solid angles for the ejecta.

The typical velocities we obtain from the simulations are similar to those calculated analytically by K16, namely around $10^4$ km/s. Furthermore, the opening angles we obtain from the simulation are similar those assumed by K16, $\Omega \approx 1 \, \rm sr$. 

\begin{figure}
\includegraphics[width=0.9\columnwidth]{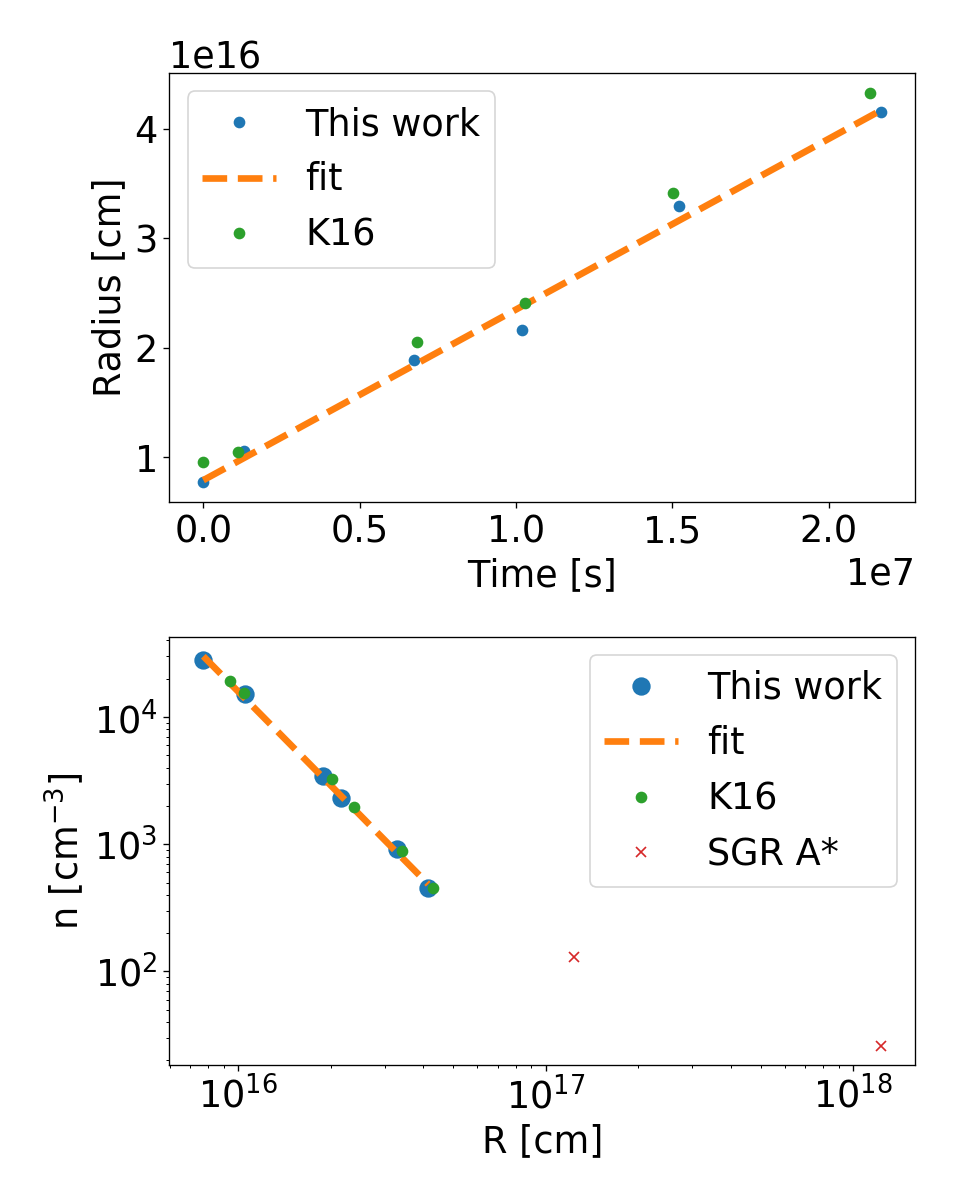}
\caption{
The shock trajectory (top panel) and the spatial density profile (bottom panel) for 14li. Results from this work are in blue; results from K16 are in green. In both panels, a power law fit to the data is shown in in orange, while in the lower panel we also include Chandra measurements of the density around our Galactic centre \citep{Baganoff2003iChandra/iGalaxy} as red crosses. Like K16, we find that the equipartition analysis yields a constant velocity of about $2 \cdot 10^4$ km/s, which is close to the value obtained in our simulations.
}
\label{fig:asassn_14li}
\end{figure}

Following the discussion in section \ref{sec:synchrotron} and using the opening angles calculated numerically (see  section \ref{sec:sim}) we use the radio observations to obtain the trajectory of the shock and the radial density profile. This is done by assuming $r = v t$ and  inverting equations \ref{eq:break_frequency} and \ref{eq:break_luminosity} to obtain the density and the velocity. This last assumption is justified because only a small faction of the ejecta's mass is involved in the shock and hence its velocity remains constant (see also K16). We assumed $\beta=7$, which implies ejecta opening angles $\theta_{y} \approx 1$ and $\theta_{z} \approx 0.1$. These results are shown in figure \ref{fig:asassn_14li}. We assumed $\chi = \epsilon_e = \epsilon_b = 0.1$, but as was pointed out in K16, the results depend very weakly on these parameters. Like K16, we find that the equipartition analysis yields a constant velocity of about $2 \cdot 10^4$ km/s, which is close to the value obtained in our simulations. The facts that the inferred velocity is constant in time and that its value is comparable to the velocity of the fastest ejecta is a consistency check for the model. A priori there was no reason that either one of these two features would arise.  
The density profile we obtained is also similar to the one obtained by K16.

The analysis here is similar, but not identical to the equipartition analysis of K16 that is also shown in Fig \ref{fig:asassn_14li}. Therefore, it is not surprising we get comparable results. The main difference is that K16 did not assume a relation between the radius and the time. Instead they assumed that the electrons responsible for the synchrotron luminosity near the break frequency dominated the total energy in electrons.  In this analysis, we do not assume anything about the relative contribution to total electron energy attributed to one part of the population, but we do assume that the radius is related to the velocity through time $r = v t$.

In these calculations we considered only  the emission from the nose of the bow shock (see figure \ref{fig:sketch}). We ignored contributions from matter that traveled downstream and from matter that enters the bow shock through the wings. In both cases the conditions differ from those at the nose.  We discuss these contributions in Appendix \ref{app:wind_emission}. It is possible that this additional emission is significant and in this case it will relax the requirement for a deep impact of the star (high $\beta$).

\section{Conclusions} \label{sec:conclusion}

Tidal disruption events produce an outflow that collides and shocks the surrounding gas, thus producing radio emission. Three mechanisms have been proposed to produce this outflow: a relativistic jet, winds from the accretion disc and the unbound stellar debris. In this work we explored the latter option. We calculated the velocity and opening angles of the fastest part of the ejecta and determined that they can account for the observed properties of the radio emission.

Using a moving mesh code \citep{Yalinewich2015Rich:Mesh} that is particularly suitable for these calculations, we simulated the tidal event itself, focusing particularly on the propagation of the unbound debris. 
We found that the characterstic speed of the ejecta is a few thousand km/s, but a small fraction of this mass expands with a larger velocity (>$10^4$ km/s). The fastest portion of the debris also subtends a larger solid angle (measured from the central black hole).   Both the velocity of the fastest-gas and and its opening angle increase with decreasing periapse distance. When this small fraction of mass collides with the diffuse ambient gas, it drives a shock wave that is several times wider than the rapidly-moving gas.  A fraction of the thermal energy in this shock wave is converted into magnetic fields and relativistic electrons that together produce synchrotron emission.

Applying these results to the very well-studied case of ASASSN-14li, we found that the opening angle and outward velocity of the debris from a tidal disruption event are consistent with those used by K16 to analyse ASASN-14li if the penetration factor of the event was large, $\beta \simeq 7$.

We assumed that the emission is dominated by shocked matter close to the nose region of the bow shock. However, it is possible that other parts of the bow shock contribute significantly and may even dominate the emission (see appendix \ref{app:wind_emission}). To the degree that the bow shock ``wings" contribute to the observed synchrotron flux, the inferred opening angle of the debris diminishes.  Because this opening angle increases with $\beta$, the result would be a true value of $\beta$ smaller than the one inferred assuming only emission from the nose region.

\citet{Kochanek1994TheStreams} and \citet{Coughlin2016OnStreams} suggested that the unbound ejecta might become gravitationally self-bound, and expand slower as a result. We note that this effect has no bearing on the radio emission, as this emission is driven by a small fraction of the mass, endowed with velocities which are considerably larger than those of the bulk.  While it is unclear that for deep penetrations a self-gravitating core would form, recent studies (Steinberg et al in preparation) have shown that deep encounters also form a self-gravitating core surrounded by high velocity non-self-gravitating gas. This faster-moving material quickly moves away from the possibly self-gravitating bulk of the ejecta, and is therefore unlikely to be significantly constrained by the debris' self-gravity.  For the same reason we do not expect this matter to clump and collapse to an even narrower wedge.

The observed radio emission enables us to use tidal disruption events to probe the density of the diffuse gas in the host galactic nuclei. In that respect, the emission from the unbound material is more useful than radio emission from a jet because the opening angle is  determined from theoretical considerations. 

Only a small fraction of tidal disruption events are thought to be radio loud \cite{Bower2012LateEvents}. Beside 14li and SWIFT J164449.3+573451 already mentioned, radio signals were also detected for two other events: XMMSL1 \citep{Alexander2017RadioJ0740-85} and ARP 299-B AT1 \citep{Mattila2018AMerger}. According to the model presented here, in those two TDEs the radio is produced by two different mechanisms (unbound debris or a jet). This might seem strange, but we argue that this is what one might expect if each of these mechanisms requires specialised conditions, and so is only effective for a small fraction of TDEs. This fraction depends on the (yet unknown) type of loss cone of other galactic centres. In the case of a full loss cone the event rate declines as $1/\beta$ \citep{Rees1988TidalGalaxies}, while for an empty loss cone the rate only declines as $1/\log \beta$ \citep{Weissbein2017HowCone}.
The magnitude of TDE radio luminosity due to the unbound debris depends on characteristics of the event (the black hole and stellar masses, the penetration factor, etc.) and also on characteristics of the environment (the interstellar density as a function of distance from the black hole).  On the other hand, radio luminosity due to a jet depends primarily on the black hole spin and the magnetic flux trapped on its event horizon \citep{Blandford1977ElectromagneticHoles,McKinney2004AHole}, as well as environmental factors \citep{vanVelzen2012ConstraintsFlares, Generozov2016TheJets}. That being said, it is worth mentioning that in many cases the radio follow - up to TDE detections were not deep or fast enough to detect a radio signal like that of 14li \citep{vanVelzen2018TheCharacterization}.

\section*{Acknowledgements}

AY would like to thank Simon Portegies Zwart, Gregg Hallinan, Chris Kochanek, Sjoert van Velzen and Nick Stone for the useful discussion. ES is supported through the NSF grant AST-1615084,  NASA  Fermi  Guest  Investigator  Program  grants NNX16AR73G  and  80NSSC17K0501;  and  through  Hubble   Space   Telescope   Guest   Investigator   Program   grantHST-AR-15041.001-A. This work made use numpy \citep{Oliphant2006ANumPy}, matplotlib \citep{Hunter2007Matplotlib:Environment} and amuse \citep{PortegiesZwart2011AMUSE:Environment}. 
This work was supported by an advanced ERC grant (TP).  It was also partially supported by NSF Grant AST-1715032 and Simons Foundation Grant 559794 (JHK).
We  acknowledge  computing  resources  from
Columbia University's Shared Research Computing Facility
project,  which  is  supported  by  NIH  Research  Facility  Improvement Grant 1G20RR030893-01, and associated funds
from the New York State Empire State Development, Division of Science Technology and Innovation (NYSTAR) Con-
tract C090171, both awarded April 15, 2010.




\bibliographystyle{mnras}
\bibliography{references}




\appendix

\section{Planar Bow Shock} \label{app:cyl_bow_shock}
Let us consider a perfectly rigid slab that extends indefinitely in the y direction and in $x>0$, and moves in the x direction with a velocity $v$ through a perfectly cold ideal gas medium. We denote the height of the slab in the z direction by $H$. As a result of this motion, a bow shock forms around the slab. We can calculate the shape of the bow shock using energy conservation. Each time interval $\Delta t$ the slab sweeps through an area $\Delta t v H$. The collision endows the gas contained in that area with specific energy (i.e. energy per unit mass) $v^2$. From this hot spot emerges a shock wave that travels in the z direction. Suppose that at time $t \gg H/v$ after the passage of the slab, the shock wave is at a distance z in the z direction. Conservation of energy dictates 
\begin{equation}
\Delta t v H \approx \Delta t v z \left(z/t\right)^2 \Rightarrow z \propto t^{2/3} \, .
\end{equation}
Since the slab is moving at a uniform velocity, the distance between the hot spot and the front of the slab is given by $x \approx v t$. Hence, very far downstream the shape of the bow shock is given by
\begin{equation}
\frac{x}{H} = C_{xz} \left(\frac{z}{H} \right)^{3/2} \, . \label{eq:bow_shock_shape}
\end{equation}
The prefactor $C_{xz}$ cannot be obtained from purely analytic considerations.

\begin{figure}
	\includegraphics[width=\columnwidth]{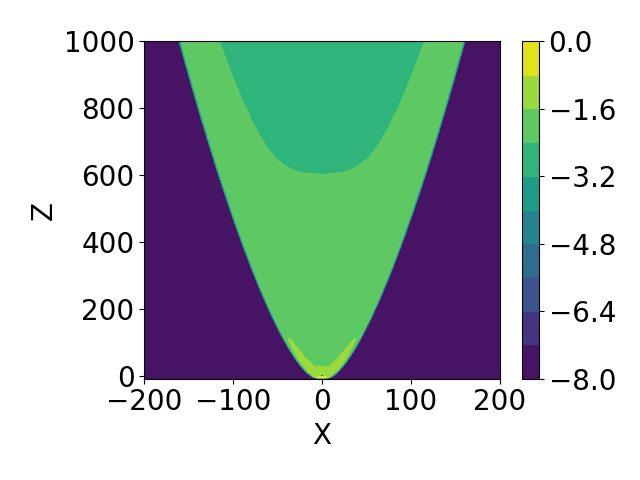}
    \caption{Pressure snapshot of the planar bow shock simulation. The x and z coordinates are measured in units of the radius of the obstacle (placed at the origin). Cold material flows from the bottom to the top.}
    \label{fig:cyl_bow_shock_snap}
\end{figure}

\begin{figure}
	\includegraphics[width=\columnwidth]{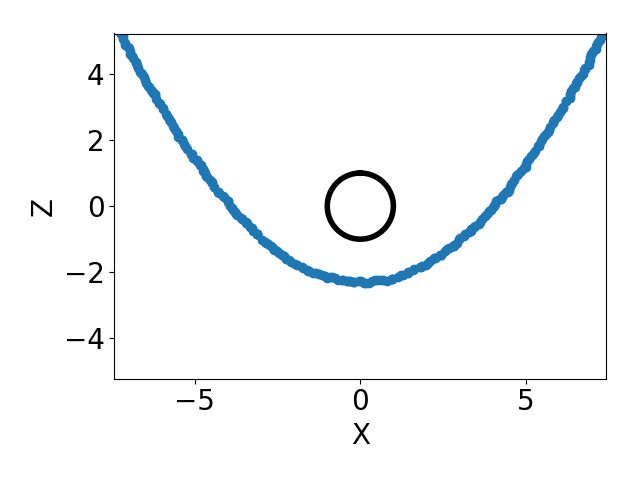}
    \caption{Zoom in on the shock front at the region closest to the nose. Axis are the same as figure \ref{fig:cyl_bow_shock_snap}. The black circle represents the obstacle.} 
    \label{fig:nose_zoom}
\end{figure}

\begin{figure}
	\includegraphics[width=\columnwidth]{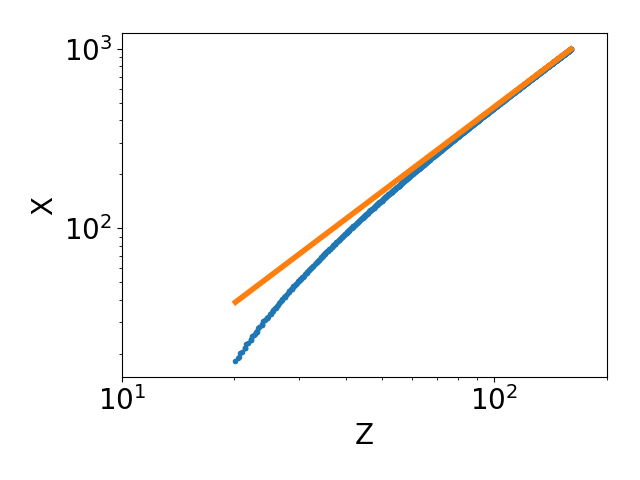}
    \caption{Shape of the shock from the simulation (blue), and a power law fit (orange). The fit agrees with the theoretical prediction (equation \ref{eq:bow_shock_shape}). All length scales are normalised to the thickness of the cylinder. The fit formula is $X = 2.9 Z^{1.57}$}
    \label{fig:cyl_bow_shock_fit}
\end{figure}

To verify equation \ref{eq:bow_shock_shape} and calibrate the coefficient $C_{xz}$, we ran a simulation using the moving mesh hydrocode RICH \citep{Yalinewich2015Rich:Mesh}. Our computational domain extends from -500 to 500 along the z direction (where length is measured in units of the radius of the cylinder), and in the x direction from -100 to 1000. The obstacle was placed such that its front is centred around the origin. Cold material (speed of sound lower by four orders of magnitude from the material velocity) flows along the positive x direction. Cells were arrange along a logarithmic spiral around the front of the slab, such that cell size inside the slab would be 0.01 (of the thickness of the slab), and the ratio between consecutive windings would be 0.01 of the radius. In this way we were able to obtain fine resolution close to the obstacle, and coarser away from it. We ran the simulation to a time ten times larger than the time it takes a fluid element to traverse the computational domain, to ensure converges to the steady state solution. A final snapshot of the pressure can be seen in figure \ref{fig:cyl_bow_shock_snap}. A zoom in near the nose of the shock front is presented in figure \ref{fig:nose_zoom}. The power law fit for the shape of the shock is shown in figure \ref{fig:cyl_bow_shock_fit}. 

\section{Emission from Bow Shock Wings} \label{app:wind_emission}

The estimates in section \ref{sec:synchrotron} consider optically thin synchrotron emission from the nose region of the bow shock (see figure \ref{fig:sketch}). We calculate here the emission from material that entered through the nose region, but travelled downstream (we call this the ``interior''), and material that entered later through the wings.

\subsection{Interior}

In this section we consider emission from material that entered the bow shock through the nose region, and travelled downstream. When the fluid element travels from the nose to a distance $x$ from the nose along the direction of the relative velocity between the obstacle and the ambient medium, its volume increases by a factor of $\left(x/H\right)^{2/3}$, where $H$ is the short axis of the obstacle. The relativistic particles cool adiabatically, so the energy of each particle scales with the volume $V$ as $V^{-1/3}$. The components of the magnetic field normal to the short axis of the obstacle diminish as $V^{-1}$, but the component parallel to the short axis remains constant. However it is possible that some of the parallel component of the magnetic field ``leaks'' onto the normal components and replenish them, and in this way this component also decays. We assume, therefore, that overall the magnetic field diminishes as $V^{-s}$, where $0<s<1$. The total emitting volume increases with depth as $\tilde{V} \propto x\cdot H \left(x/H\right)^{2/3}$, but the density decreases as $V^{-1}$. Putting all this together, we get that the optically thin synchrotron luminosity at a given frequency scales with depth as

\begin{equation}
\frac{d \ln L_{\omega}}{d \ln x} = - \frac{p s}{3} + \frac{2 p}{9} - \frac{s}{3} + \frac{7}{9}
\end{equation}
Since $1>s>0$ and $3>p>2$

\begin{equation}
\frac{13}{9} > \frac{d \ln L_{\omega}}{d \ln x}>\frac{1}{9}
\end{equation}
So the emission is always dominated by the deepest (i.e. farthest from the nose) parts of the bow shock. However, if indeed $s=1$, then this dependence is very weak, so all depths contribute similarly. The contribution from the deepest layers will be larger than that of the nose by a factor of $\theta_{z}^{-d \ln L_{\omega}/ d\ln x}$. In the case of large values of $\frac{d \ln L_{\omega}}{d \ln x}$, the luminosity can be considerably larger than calculated in section \ref{sec:synchrotron}.

\subsection{Wing}

For material that entered the bow shock from the wings. The shock becomes weaker because the normal component of the velocity decreases with depth as $v_{\perp} \propto \left(x/H\right)^{-1/3}$. The optically thin synchrotron luminosity at a given frequency scales with depth as

\begin{equation}
\frac{d \ln L_{\omega}}{d \ln x} = \frac{13 - 5 p}{6}
\end{equation}
When $p<\frac{13}{5} = 2.6$ the emission is dominated by the deepest layer, but if $p>2.6$ then the nose is always brighter than the wings. In the former case, the luminosity will be considerably larger than what was calculated in section \ref{sec:synchrotron}.

In the case when the luminosity is dominated by the deepest layers, and the deepest layer is at a distance $r$ from the nose, then according to section \ref{app:cyl_bow_shock} the width of the bow shock along the z axis is $r \theta_z^{1/3}$, and the normal component is smaller than the absolute magnitude of the velocity by a factor of $\theta_z$. The break frequency in this case is given by
\begin{equation}
    \omega_{\rm sa} \approx c^{\frac{2 \left(- 2 p + 3\right)}{p + 4}} \chi^{\frac{2 \left(- p + 2\right)}{p + 4}} \epsilon_{b}^{\frac{p + 2}{2 \left(p + 4\right)}} \epsilon_{e}^{\frac{2 \left(p - 1\right)}{p + 4}} m_{e}^{\frac{- 5 p + 2}{2 \left(p + 4\right)}} r^{\frac{2}{p + 4}}  \times
\end{equation}
\begin{equation*}
    \theta_{z}^{\frac{5 p + 4}{3 \left(p + 4\right)}} \left(\sqrt{m_{p}} v\right)^{\frac{5 p - 2}{p + 4}} \left(\sqrt{n} \sqrt{r_{e}}\right)^{\frac{p + 6}{p + 4}}
\end{equation*}
and the break luminosity si given by
\begin{equation}
    L_{\omega, \rm sa} \approx \chi^{\frac{5 \left(- p + 2\right)}{p + 4}} \epsilon_{b}^{\frac{2 p + 3}{2 \left(p + 4\right)}} \epsilon_{e}^{\frac{5 \left(p - 1\right)}{p + 4}} \theta_{y} \theta_{z}^{\frac{4 \left(3 p + 2\right)}{3 \left(p + 4\right)}} \times
\end{equation}
\begin{equation*}
     \left(c^{5} m_{e}^{\frac{5}{2}}\right)^{\frac{- 2 p + 3}{p + 4}} \left(\sqrt{m_{p}} v\right)^{\frac{12 p - 7}{p + 4}} \left(\sqrt{n} r \sqrt{r_{e}}\right)^{\frac{2 p + 13}{p + 4}}
\end{equation*}
in contrast to the expression obtained in section \ref{sec:synchrotron}.
 

\bsp	
\label{lastpage}
\end{document}